\pdfoutput=1 

\documentclass[12pt,a4paper]{iopart}
\usepackage{iopams,setstack}
\usepackage{array}
\usepackage{multirow}
\usepackage{graphicx}

\usepackage[colorlinks, linkcolor=blue, citecolor=blue, urlcolor=blue, breaklinks=true]{hyperref}

\renewcommand{\eref}[1]{Eq.~(\ref{#1})}
\renewcommand{\Eref}[1]{Equation~(\ref{#1})}
\renewcommand{\fref}[1]{Fig.~\ref{#1}}
\renewcommand{\sref}[1]{Sec.~\ref{#1}}
\renewcommand{\tref}[1]{Table~\ref{#1}}

\renewcommand{\rmd}{\mathrm{d}}
\renewcommand{\Tr}{\mathrm{Tr}}

\usepackage{etoolbox}

\usepackage{lipsum}

\makeatletter
\def\@mkboth#1#2{}
\newlength\appendixwidth
\preto\appendix{\addtocontents{toc}{\protect\patchl@section}}
\newcommand{\patchl@section}{%
  \settowidth{\appendixwidth}{\textbf{Appendix }}%
  \addtolength{\appendixwidth}{1.5em}%
  \patchcmd{\l@section}{1.5em}{\appendixwidth}{}{\ddt}%
}
\makeatother

\begin{document}

\title{Thermodynamics of trajectories of open quantum systems, step by step}

\author{Simon Pigeon}
\address{Centre for Theoretical Atomic, Molecular and Optical Physics, School of Mathematics and Physics, Queen's University Belfast, Belfast BT7\,1NN, United Kingdom}
\ead{s.pigeon@qub.ac.uk}

\author{Andr\'e Xuereb}
\address{Department of Physics, University of Malta, Msida MSD\,2080, Malta}
\address{Centre for Theoretical Atomic, Molecular and Optical Physics, School of Mathematics and Physics, Queen's University Belfast, Belfast BT7\,1NN, United Kingdom}
\ead{andre.xuereb@um.edu.mt}


\begin{abstract}
Thermodynamics of trajectories promises {to make possible the} thorough analysis of the dynamical properties of an open quantum system, a sought-after goal in modern physics. Unfortunately, calculation of the relevant quantities presents severe challenges. Determining the large-deviation function that gives access to the full counting statistics associated with a dynamical order parameter is challenging, if not impossible, even for systems evolving in a restricted Liouville space. Acting on the realisation that the salient features of most dynamical systems are encoded in the first few moments of the counting statistics, in this article we present a method that gives sequential access to these moments. Our method allows for obtaining analytical result in several cases, as we illustrate, and allows using large deviation theory to reinterpret certain well-known results.
\end{abstract}

\pacs{
03.65.Aa
,03.65.Yz
,42.50.Lc
}
\vspace{2pc}
\noindent{\it Keywords}: dissipative systems, exact results, large deviations in non-equilibrium systems

\tableofcontents

\section{Introduction}

Thermodynamics of trajectories proposes a set of powerful tools that fully characterises the dynamics of open systems~\cite{Touchette2009}. The formalism was only recently extended from classical to quantum systems~\cite{Garrahan2010}, promising to shed a new light on the {behaviour} of even well-known quantum systems. Through the determination of the scaled cumulant generating function, also known as the large-deviation function, the full counting statistics of the exchange of excitations taking place between a system and its environment can be studied. This approach allows {one} to spot dynamical phase transitions directly through the determination of the large-deviation function, promising it a bright future~\cite{Lesanovsky2013}.

Despite the fact that this large-deviation approach is founded on simple principles, however, severe difficulties are generally encountered when applying it to physical systems. Chief amongst these is the difficulty in determining the large-deviation function; indeed, only a few analytical results are known from a handful of classical and quantum systems~\cite{Touchette2009,Derrida1998,Pigeon2014a,Saito2007,Znidaric2014}. Naturally, this dramatically restricts the usefulness of applying thermodynamics of trajectories and makes the use of dedicated numerical methods essential~\cite{Giardina2006,Nemoto2014}.

In this paper we present an alternative route towards accessing the same information that is encoded in the large-deviation function, by exploiting the fact that the first few cumulants of the probability distribution of the exchange statistics already contain a wealth of information about this distribution. We obtain an efficient iterative approach that allows one to determine successive cumulants, giving access to {characteristics of} the dynamics that are usually not easily accessible or calculated. Furthermore, this method allows one to gain a deeper understanding of certain features in the dynamics of open quantum systems, such as dynamical phase transitions. It also allows {one} to reinterpret well-known phenomena in open quantum systems, as we illustrate {here} by studying an optical system that presents bistability. {The method we propose here has the added benefit of being conceptually similar to approaches used previously that were tailored for specific phenomena, such as resonance fluorescence~\cite{Lenstra1982,Sanchez2007} and electronic transport statistics~\cite{Flindt2008,Flindt2010}.}

In what follows, we begin {by introducing} thermodynamics of trajectories of open quantum systems and detailing the iterative method we propose. Next, we start exploring the application of our method by using it to study a small, two-qubit, quantum system; this reveals the usefulness of our approach even in the case where alternative approaches are available. Finally, we {study} a system associated with an infinite Liouville space, i.e., a quantum harmonic oscillator, for which the usual non-numerical approaches to determining the large-deviation function fail.

\section{Full-counting statistics without {a} large-deviation function}
We devote this section to a formal development of the proposed method. First, we briefly introduce thermodynamics of trajectories as a formalism that helps {to analyse} the dynamics of open quantum systems, and as developed in Ref.~\cite{Garrahan2010}. Following this, we describe an iterative method that allows us to access successive cumulants of the statistics of exchange of excitations between the system and its environment. Finally, we focus on applying this method to the analysis of a situation presenting dynamical multi-stability, and we explore signatures of multi-stability in the cumulants obtained.

\subsection{Thermodynamics of trajectories for open quantum system}
The method we explore is applicable to any quantum system, {as} described by its density matrix $\rho$, which undergoes a dynamical evolution governed by the master equation $\partial_{t}\rho=\mathcal{W}\left[\rho\right]$. The superoperator $\mathcal{W}$, which drives the evolution, can be decomposed {into} a Hamiltonian contribution and a dissipative part, viz.
\begin{equation}
\mathcal{W}[\bullet]=-i\bigl[\hat{H},\bullet\bigr]+\mathcal{L}[\bullet].\label{eq:master}
\end{equation}
In the case of Markovian systems, the Lindbladian superoperator $\mathcal{L}$ can be {further} decomposed in terms
of a family of so-called Liouville operators $\hat{L}_i$ acting on the system, together with associated rates $\gamma_i\geq0$, in the following way:
\begin{equation}
\mathcal{L}[\bullet]=\sum_i\gamma_i\left(\hat{L}_i\bullet\hat{L}_i^\dagger-\frac{1}{2}\bigl\{ \hat{L}_i^\dagger\hat{L}_i,\bullet\bigr\}\right),
\end{equation}
where each index $i$ denotes a different \emph{channel} connecting the system to its environment. In the following, we may drop the index if the system has only one dissipation channel. We consider a single one of these channels, {with} which we associate a counting process $K:=K_i$; more details regarding this can be found in Ref.~\cite{Garrahan2010}. We can now define the moment generating function, which gives access to all the moments (equivalently, cumulants) of the probability distribution associated with the statistics of $K$. To this end, we apply large-deviation theory~\cite{Touchette2009} and write
\begin{equation}
Z(s,t)=\Tr\{\varrho_{s}\}\underset{t\to\infty}{\longrightarrow}e^{t\theta(s)},
\end{equation}
where $\varrho_{s}(t)=\sum_{K}e^{-sK}P^{K}\rho(t)$, with $P^{K}$ being the projector over the subspace corresponding to a count $K$, is called the {\emph{biased density matrix}. It can be decomposed into sets of trajectories named \emph{$s$-ensemble} or \emph{biased-ensemble} whose weights differ from the probabilities inherent in the density matrix $\rho(t)$ by a factor exponential in the respective count $K$}. In the long-time limit, this partition {function takes the so-called} large-deviation form shown in the preceding equation, where the function $\theta(s)$ in the exponent is the scaled cumulant generating function, also known as the large-deviation function, with $s$ {the} parameter conjugate the the counting process $K$. For simplicity we restrict ourselves in this article to counting processes associated with a single channel. Nevertheless, the very same framework described here and in the next section can be derived for other, more general, counting processes.

The $s$-biased ensemble, which is unnormalised and therefore cannot be interpreted as a density matrix, undergoes a dynamics governed by the equation
\begin{equation}
\partial_{t}\varrho_{s}=\mathcal{W}[\varrho_{s}]+\mathcal{L}_{s}[\varrho_{s}],\label{eq:master-biased}
\end{equation}
where the biasing superoperator $\mathcal{L}_s$ is given by
\begin{equation}
\mathcal{L}_{s}[\bullet]=\gamma_i(e^{-s}-1)\hat{L}_i\bullet\hat{L}_i^\dagger.\label{eq:b-superop}
\end{equation}

{Taking the time derivative of the cumulant generating function $\ln Z(s,t)$  we find that $\partial_t \ln Z(s,t)=\Tr \left\{ \mathcal{L}_{s}[\varrho_{s}] \right\}/\Tr \left\{ \varrho_{s} \right\}$, because the superoperator $\mathcal{W}$ is trace-preserving. Consequently, using the definition of the biased superoperator $\mathcal{L}_s$ as given in \eref{eq:b-superop}, t}he large-deviation function can be evaluated using~\cite{Ates2012}
\begin{equation}
\theta(s)=\gamma_i(e^{-s}-1)\lim_{t\to\infty}\frac{1}{t}\int_0^t\rmd\tau\,\Tr\bigl\{\hat{L}_i^\dagger\hat{L}_i\rho_{s}(\tau)\bigr\},\label{eq:ldf-integral}
\end{equation}
where we define a {normalised biased} density matrix through the relation $\rho_{s}:=\varrho_{s}/\Tr\{\varrho_{s}\}$, whose evolution is governed by the biased master equation
\begin{equation}
\partial_{t}\rho_{s}=\mathcal{W}[\rho_{s}]+\mathcal{L}_{s}[\rho_{s}]-\rho_{s}\Tr\{\mathcal{L}_{s}[\rho_{s}]\}. \label{eq:s}
\end{equation}
\Eref{eq:ldf-integral} implies that the large-deviation function $\theta(s)$ can be evaluated as an integral over time of an observable of the biased density matrix $\rho_s$, and \emph{not} of the ``standard'' density matrix $\rho$. Consequently, the rich information contained in the large-deviation function and related to the dynamics of the system cannot be directly accessed through $\rho$, but only through $\rho_s$. More concretely, $\theta(s)$ encodes the full counting statistics of the counting process $K$ through the family of relations
\begin{equation}
\partial_s^n\theta(s)\big\vert_{s=0}=(-1)^{n}\kappa_n,\label{eq:moment}
\end{equation}
where $\kappa_n$ is the $n$\textsuperscript{th} cumulant of the probability distribution of the counting process being considered. Typically this procedure can be used to, e.g., identify regimes in which different dynamical phases coexist and reveal dynamical phase transitions between such phases~\cite{Garrahan2007,Hickey2012,Hickey2013,Pigeon2014b}. The presence of distinct dynamical phases in a system can be related to {a} non-analyticity at $s=0$ of the large-deviation function $\theta(s)$. We will return to this point in \sref{sub:Multi-stability-and-dynamical}, where we shall connect it to the proposed approach. Other key thermodynamics statements, such as fluctuation relations~\cite{Evans2002,Pigeon2015}, may be made once the large-deviation is known.

As discussed in the introduction, despite the well-known usefulness of the large-deviation function following from the formalism of thermodynamics of trajectories~\cite{Touchette2009,Lecomte2007} as described above, its determination appears to be very challenging, both in the classical and quantum regimes. Indeed, dedicated numerical methods have been developed for this {purpose}~\cite{Giardina2006,Nemoto2014}. In the present context, the difficulty presents itself in evaluating the long-time {behaviour} of the biased density matrix $\rho_s$ by means of \eref{eq:ldf-integral}. Explicitly solving this equation, or the evolution of specific observables, appears to be difficult for most systems{;} the only possibility in most cases is numerical evaluation. In the following we illustrate how one can access exactly and analytically the cumulants $\kappa_n$, the first few of which contain the majority of the information of interest regarding the dynamics in question, by means of an iterative approach.

\subsection{Iterative evaluation of the cumulants\label{sub:Iterative}}
By using the definition of $\theta(s)$ given in \eref{eq:ldf-integral} and connecting it to the cumulants by means of the equation in \eref{eq:moment} we have for the first cumulant $\kappa_1$ the following straightforward
relation:
\begin{equation}
\kappa_1=\gamma_i\lim_{t\to\infty}\frac{1}{t}\int_0^t\rmd\tau\Tr\bigl\{\hat{L}_i^\dagger\hat{L}_i\rho(\tau)\bigr\},\label{eq:11}
\end{equation}
which links $\kappa_1$ to the long-time evolution of the expectation value of operators with respect to the original density matrix $\rho$. We see immediately that the first cumulant may be evaluated from observables of the original dynamics, even if the full {statistics} of the exchange may not. This being so, we therefore conclude that if one is interested only in the the first cumulant, $\kappa_1$, which is quite often experimentally the one most easily accessed, the difficulty is reduced from evaluating the long-time {behaviour} of $\rho_s(t)$ to evaluating that of $\rho(t)$. In the case of systems converging to a steady state, one need only concern themselves with the steady-state density matrix $\tilde{\rho}$, which is the solution of the steady-state equation $\mathcal{W}[\tilde{\rho}]=0$. By contrast, obtaining the large-deviation function requires determining the steady-state $\tilde{\rho}_s$, associated with the biased density matrix $\rho_s(t)$ that solves \eref{eq:s}, which by construction depends on the bias parameter $s$ and obeys a more complex evolution.

Let us now take a step further and investigate the second cumulant, and moment, of the distribution. Once again making use of \eref{eq:moment}, we find
\begin{eqnarray}
\kappa_2=\gamma_i\lim_{t\to\infty}\frac{1}{t}\int_0^t\rmd\tau & \Bigl[\Tr\{ \hat{L}_i^\dagger\hat{L}_i\rho(\tau)\} \nonumber \\
& -2\Tr\{ \hat{L}_i^\dagger\hat{L}_i\rho'(\tau)\}\Bigr],\label{eq:k2}
\end{eqnarray}
with $\rho'=\partial_{s}\rho_{s}\bigr|_{s=0}$, which we term the first-order biased matrix; higher orders will be defined analogously as higher derivatives of $\rho_s$ with respect to the bias parameter $s$. The matrix $\rho'$ evolves according to the bare (or unbiased) system dynamics, governed by the superoperator $\mathcal{W}$, plus a component related to the biasing procedure given in \eref{eq:b-superop}:
\begin{equation}
\partial_t\rho'=\mathcal{W}[\rho']-\gamma_i\hat{L}_i\rho\hat{L}_i^\dagger+\gamma_i\rho\Tr\{\hat{L}_i^\dagger\hat{L}_i\rho\}.\label{eq:rho'me}
\end{equation}
We note that, since $\kappa_2$ depends on observables of $\rho'$, it cannot be obtained merely through observables of $\rho$ itself. It is also possible to show that $\rho'$ is traceless; indeed, $\Tr\{\rho'\}=\Tr\{\varrho'-\rho\Tr\{\varrho'\}\}$, where $\varrho'=\partial_{s}\varrho_{s}\bigr|_{s=0}$ and the $s$-biased matrix $\rho_s$ is defined above. The equation governing the evolution of $\rho'$ also tells us that in order to calculate how $\rho'$ evolves in time we first need to determine the evolution of the unbiased density matrix $\rho$.

We may in fact proceed further in exactly the same way and obtain an analogous expression for the third cumulant:
\begin{eqnarray}
\kappa_3=\gamma_i\lim_{t\to\infty}\frac{1}{t}\int_0^t\rmd\tau\Bigl[&\Tr\{\hat{L}_i^\dagger\hat{L}_i\rho(\tau)\}-3\Tr\{\hat{L}_i^\dagger\hat{L}_i\rho'(\tau)\}\nonumber\\
&\qquad+3\Tr\{\hat{L}_i^\dagger\hat{L}_i\rho''(\tau)\}\Bigr],
\end{eqnarray}
with $\rho''=\partial_{s}^2\rho_{s}\bigr|_{s=0}$ is the second-order biased matrix, which now depends on the unbiased density matrix $\rho$ as well as the first-order biased matrix $\rho'$, which {in turn} undergoes a dynamics governed by \eref{eq:rho'me}. For $\rho''$ we find
\begin{eqnarray}
\partial_t\rho''=\mathcal{W}[\rho'']+& 2\gamma_i\rho'\Tr\{\hat{L}_i^\dagger\hat{L}_i\rho\}+\gamma_i\hat{L}_i(\rho-2\rho')\hat{L}_i^\dagger \nonumber\\
& -\gamma_i\rho\Tr\{\hat{L}_i^\dagger\hat{L}_i(\rho-2\rho')\}.
\end{eqnarray}
We notice once again that the second-order biased matrix is traceless, i.e., $\Tr\{\rho''\}=0$. As per that of $\rho'$, the evolution of $\rho''$ is related to the bare dynamics through $\mathcal{W}$, this time with a biasing part related to $\rho$ and $\rho'$.

This procedure may be iterated further in order to obtain expressions for successive cumulants; in order to determine the $n$\textsuperscript{th} cumulant, one needs to solve the evolution of $n$ matrices ($\rho$, $\rho'$, and so on up to $\rho^{(n)}$). More directly, we can rephrase our method in terms of the expectation values $\langle\bullet\rangle_j:=\Tr\{\bullet\rho^{(j)}\}$ ($j=0,1,2,\dots$): In order to calculate $\kappa_n$, one needs to solve for $\langle\hat{L}_i^\dagger\hat{L}_i\rangle_j$ and $\langle\hat{L}_i^{\dagger2}\hat{L}_i^2\rangle_j$ for each $j<n$, where we remind that the index $i$ refers to the channel under investigation. We may in fact explicitly write
\begin{eqnarray}
\kappa_n&=\left(-1\right)^n\lim_{t\to\infty}\frac{1}{t}\int_0^t\rmd\tau\Tr\{\mathcal{L}^{(n)}[\rho(\tau)]\}\\
&=\gamma_i\sum_{j=0}^{n-1}{{n}\choose{j}}(-1)^j\lim_{t\to\infty}\frac{1}{t}\int_0^t\rmd\tau\Tr\{\hat{L}_i^\dagger\hat{L}_i\rho^{(j)}(\tau)\},
\end{eqnarray}
where we have made use of the definition
\begin{eqnarray}
\mathcal{L}^{(n)}[\rho]&:=\partial_s^n\mathcal{L}_s[\rho_{s}]\bigr|_{s=0}\\
&=\gamma_i\sum_{j=0}^{n-1}{{n}\choose{j}}(-1)^{n-j}\hat{L}_i\rho^{(j)}\hat{L}_i^\dagger.
\end{eqnarray}
Extending the preceding work, we find that the $n$\textsuperscript{th}-order biased matrix $\rho^{(n)}=\partial_{s}^{n}\rho_{s}\big\vert_{s=0}$ evolves according to the equation
\begin{eqnarray}
\partial_t\rho^{(n)}&=\mathcal{W}[\rho^{(n)}]+\mathcal{L}^{(n)}[\rho]\nonumber\\
&\qquad-\sum_{j=0}^{n-1}{{n}\choose{j}}\rho^{(j)}\Tr\{\mathcal{L}^{(n-j)}[\rho]\}\label{eq:ccdcd}\\
&=\mathcal{W}[\rho^{(n)}]+\gamma_i\sum_{j=0}^{n-1}{{n}\choose{j}}\Biggl[(-1)^{n-j}\hat{L}_i\rho^{(j)}\hat{L}_i^\dagger\nonumber\\
&\quad-\sum_{k=0}^{j-1}{{j}\choose{k}}(-1)^{j-k}\Tr\{\hat{L}_i^\dagger\hat{L}_i\rho^{(k)}\}\rho^{(n-j)}\Biggr],
\end{eqnarray}
where $\Tr\{\rho^{(n)}\}=\delta_{n,0}$. The first of the preceding two equations{, \eref{eq:ccdcd},} corresponds to the generic definition, which can be used to explore counting processes other than the one being considered here and depends on different orders of $\mathcal{L}^{(n)}$. The second is specialised for the biased superoperator defined in \eref{eq:b-superop}. When the system converges to a time-independent steady state, the definitions above simplify, since $\lim_{t\to\infty}\frac{1}{t}\int_0^t\rmd\tau\Tr\{\rho^{(n)}(\tau)\}\to\Tr\{\tilde{\rho}^{(n)}\}$, where $\tilde{\rho}^{(n)}$ is the stationary solution of $\rho^{(n)}(t)$ for the dynamics in question. In this section, we have described the promised method mostly in the language of density matrices. In \ref{sec:a1} the same method is described through expectation values computed on the different orders of the biased matrix. Depending on the nature of the specific system under consideration, especially if its Liouville space is large, this alternative may prove more useful than the matrix-based method as discussed above.

By way of summary, the essential idea underlying our method is that once the first $n-1$ cumulants of the counting statistics are computed, it is relatively straightforward to extend the calculation to the $n$\textsuperscript{th} cumulant. This method avoids a numerical coarse-graining of the large-deviation function $\theta(s)$ and also allows a close comparison to be made to experimental data, since in such data it is only the first cumulants (or moments) that are accessible, rather than the full counting statistics. To elaborate on this point, we shall now focus on using the language of thermodynamics of trajectories to define dynamical phase transitions and draw a connection between this dynamical feature and the present method. {We note here that Ref.~\cite{Flindt2013} presents work, similar in spirit to ours, that focuses on higher-order cumulants.}

\subsection{Multi-stability and dynamical phase transitions\label{sub:Multi-stability-and-dynamical}}
We have already discussed how the large-deviation theory presents itself as a powerful tool to analyze the dynamics of a system. It is known that the occurrence of phenomena such as dynamical phase transitions can be identified through the large-deviation function $\theta(s)$~\cite{Garrahan2007,Hickey2012,Hickey2013,Pigeon2014b}. In this context, the large-deviation function behaves as a free energy function, with $s$ playing the role of temperature. Non-analyticities at $s=0$ indicate that the system being explored can converge, in the long-time limit, towards states for which the counting process displays distinctly different {behaviour}. In other words, in situations where the dynamics converges to a steady state, a non-analyticity in $\theta(s)$ at $s=0$ {identifies} multiple fixed points for the dynamical map corresponding to the evolution of the system. We will discuss how the iterative method we present here may give access to this information without requiring one to evaluate the large-deviation function itself. This enables one to identify dynamical phase transitions even in situations where the large-deviation function cannot be calculated easily or at all, and consequently it extends the range of applicability of thermodynamics of trajectories. Before continuing, we note that $\theta(s)$ may also display non-analyticities at $s\neq0$~\cite{Touchette2009}, corresponding to more complex {behaviour} where dynamically distinct {behaviour} may be observed in the tails of the probability distribution corresponding to the counting process. Such {behaviour} cannot be revealed through the method we present.

A non-analyticity in the large-deviation function $\theta(s)$ at $s=0$ translates into two distinct values for at least one cumulant, say the $n$\textsuperscript{th}, as $s\to0$:
\begin{equation}
\kappa_n^\pm=(-1)^n\lim_{s\to0^\pm}\partial_{s}^{n}\theta(s);
\end{equation}
where $\kappa_n^+\neq\kappa_n^-$. We note that it is common for $\kappa_n^+\neq\kappa_n^-$ to imply that $\kappa_j^+\neq\kappa_j^-$ for $j>n$, but this is not necessary. The fact that the $n$\textsuperscript{th} cumulant has two values as $s\to0$ signals the presence of at least two distinct dynamical {behaviours}, i.e., distinct fixed points of the dynamical map. We may say that a dynamical phase transition is of order $n$ if $\kappa_j^+=\kappa_j^-$ for $j<n$ but $\kappa_n^+\neq\kappa_n^-$. The role played by the order parameter in micro-canonical phase transitions is taken up in dynamical phase transitions by the counting process $K$~\cite{Lesanovsky2013}.

To understand how the framework we present can be used advantageously in this context, let us consider the case of a first-order dynamical phase transition, where $\kappa_1^+\neq\kappa_1^-$. These two distinct first cumulants reflect that the system converges dynamically toward at least two steady states, such that
\begin{equation}
\kappa_1^{\pm}=\gamma_i\Tr\{\hat{L}_i^\dagger\hat{L}_i\tilde{\rho}_\pm\},\label{eq:cddc-1-1}
\end{equation}
where $\tilde{\rho}_\pm$ being two solutions of the equation $\mathcal{W}[\tilde{\rho}_\pm]=0$. As discussed previously, the first cumulant depends exclusively on the unbiased dynamics. Consequently, from here we can infer that any quantum open system presenting at least two {fixed points (or attractors)} in its dynamical map {with distinct dynamical behaviours} will have two possible values for the first cumulant, as shown in \eref{eq:cddc-1-1}, and therefore present a dynamical phase transition. 

In the present scenario, if the dynamical map possesses more than two attractors, the first cumulant $\kappa_1$ will be still only dually defined. {This originates directly from the thermodynamics of trajectories approach, which focuses on the {behaviour} of the tails of the probability distribution $p_K(t)=\Tr\{P^K\rho(t)\}$. Only the attractors dominating the tails of the distribution towards the lower- ($s\to0^-$) or upper- ($s\to0^+$) limit of the counting process $K$~\cite{Touchette2009} will have an effect on the large deviation function.} In other words, use of all solutions of $\mathcal{W}[\tilde{\rho}]=0$ with the method we discuss here gives more information than even the full determination of the large-deviation function, as we shall see in the next section for a system presenting a dynamical phase transition. Despite the wealth of examples of first-order dynamical phase transitions that are known, examples of higher-order dynamical phase transitions are {not known to us.} 

We shall now proceed to apply our formalism to a number of paradigmatic examples. We will first consider a system composed of two spins-$1/2$, which evolves in a small Liouville space and in which case we can easily compare the results we obtain by applying the framework we discuss here to the direct determination of the large-deviation function. We will then turn our attention to a system composed of harmonic oscillators, which evolves within an infinite-dimensional Liouville space and for which $\theta(s)$ cannot be determined directly. For both of these examples we will first consider one or more systems that present only a single dynamical phase, using this to illustrate how the method proposed here gives analytical results that are not accessible through the numerical determination of $\theta(s)$. Following this, we will discuss an example of a dynamical phase transition, where the dynamics converges to two distinct steady states.

\section{Damped spin systems}
Spin systems, especially ones composed of spins-$1/2$, have the strong advantage in the present context of residing on a rather small Liouville space: The dynamics of such a system may easily be resolved through analytic or numerical methods. For what concerns us, the full thermodynamics of trajectories of such a system can be accessed directly by studying the superoperator $\mathcal{W}_s:=\mathcal{W}+\mathcal{L}_s$ presented in \eref{eq:master-biased}. For cases where the long-time dynamics of the density matrix $\rho_s(t)$ cannot be directly resolved in matrix form, it is still possible to vectorise the biased master equation and to obtain $\theta(s)$ directly as the slowest-decaying eigenvalue of the matrix corresponding to $\mathcal{W}_s$, i.e., the eigenvalue with the most positive real part. Both approaches are equivalent and give direct access to the large-deviation function, thereby permitting a full characterisation of the dynamics of the system through the counting process $K$. On the one hand, for any but the simplest of systems, this very powerful method reduces to numerical diagonalisation of matrices, and therefore cannot be used to derive analytic expressions of the cumulants of the counting process. On the other hand, the iterative method we propose here does give access to these analytical expressions. We begin our illustrations of this method by considering a system composed of two spins-$1/2$ individually damped by a zero-temperature environment.

\subsection{Two individually damped spins-$1/2$: Same polarisability}
Consider a system composed of two spins-$1/2$ evolving under the Hamiltonian
\begin{equation}
\hat{H}=\hat{\sigma}_1^\mathrm{x}\hat{\sigma}_2^\mathrm{x}+h(\hat{\sigma}_1^\mathrm{z}+\hat{\sigma}_2^\mathrm{z}),\label{eq:spin1}
\end{equation}
where $\hat{\sigma}_i^\mathrm{x,y,z}$ ($i=1,2$) are the Pauli operators operating on spin $i$ and $h$ the strength of an externally-applied magnetic field. We assume that both spins are damped individually by means of the Liouville operators $\hat{L}_i=\hat{\sigma}_i^+$ ($i=1,2$); for simplicity, we take the damping rates to be equal, i.e., $\gamma:=\gamma_1=\gamma_2$. Despite its simplicity, this system does not allow for a straightforward analytical determination of its large-deviation function. By contrast, the method we propose gives access to analytical expressions to be obtained for successive cumulants of the spin-flip statistics.

Let us consider as a counting process the spin-flips of one of the two spins, say the the first ($K:=K_1$). {In order to determine the first cumulant we must first determine the steady-state solution of the density matrix of the system $\rho$. Subsequently, using \eref{eq:11}, we find the first cumulant
\begin{equation}
\kappa_1=\frac{2\gamma}{\gamma^2+16h^2+4}.\label{eq:mean1}
\end{equation}
Based on the steady-state solution of the density matrix, we can determine the steady-state solution of the first-order biased density matrix $\rho'$ according to \eref{eq:rho'me}, which then allows us to obtain the second cumulant, $\kappa_2$, as defined in \eref{eq:k2}. The corresponding Fano factor can then be calculated and found to be given by}
\begin{equation}
\frac{\kappa_2+\kappa_1^2}{\kappa_1}=1+\frac{32(4h^2+1)}{(16h^2+\gamma^2+4)^2}+\frac{2(\gamma-5)}{16h^2+\gamma^2+4}+\frac{2}{\gamma^2+4}.\label{eq:fano1}
\end{equation}

\begin{figure}
\centering
\includegraphics[scale=0.55]{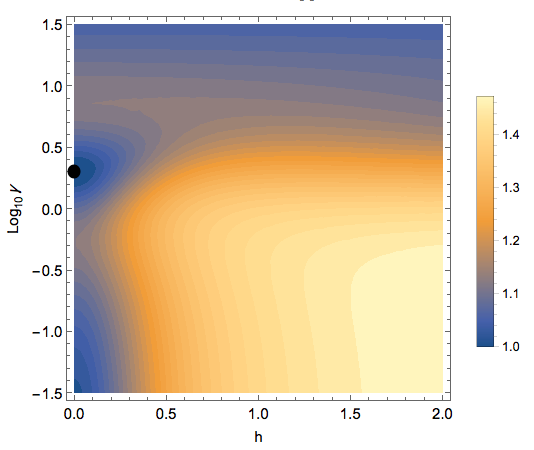}
\caption{Fano factor of the {spin-flip} statistics of one of two interacting spins-$1/2$. The plot is given as a function of the magnetic field $h$ on the horizontal axis and the spin flip rate $\gamma$ on the vertical. {At zero magnetic field ($h=0$), a minimum is found for $ \gamma =  2 $ (black dot).} \label{fig:XZfano}}
\end{figure}

In \fref{fig:XZfano} we represent the Fano factor as a function of the magnetic field $h$ and the spin-flip rate $\gamma$. A feature to notice is that for {strong damping}, the spin-flip statistics converge to a Poissonian distribution, i.e., $(\kappa_2+\kappa_1^2)/\kappa_1\to1$. {We note that for zero magnetic field ($h=0$) and finite $\gamma$ the exchange statistics is Poissonian only when $\gamma=2$}. We see here that despite having numerical access to the full-counting statistics we can use the present approach in order to efficiently and exactly evaluate the most relevant moments of the exchange statistics, thereby retrieving rich information whilst bypassing any need for numerical calculations.

From this example we see here that our iterative approach is extremely valuable for evaluating the parameters of a system from the statistics of its output. In our next example, we will be focus particularly on this aspect.

\subsection{Two {individually-damped} spins-$1/2$: Inverse polarisability}
\begin{figure}
\centering
\includegraphics[scale=0.55]{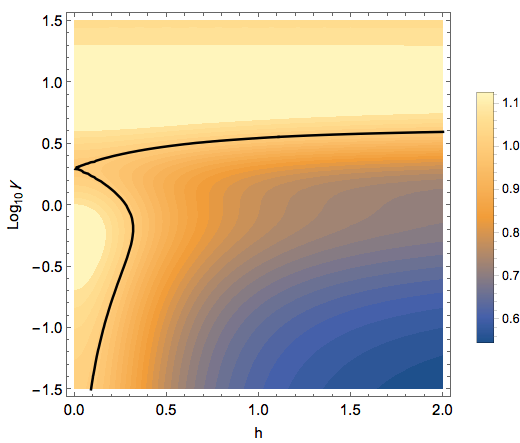}
\caption{Fano factor of the {spin-flip} statistics of one of two interacting spins-$1/2$. The plot is given as a function of the magnetic field $h$ on the horizontal axis and the spin flip rate $\gamma$ on the vertical. The solid black line corresponds to a Poissonian distribution; the lower and upper regions delineated by this line correspond respectively to antibunching (sub-Poissonian) and bunching statistics.\label{fig:XZfano-1}}
\end{figure}

Consider a system identical to the one in the previous section, but for the magnetic susceptibility of spin $2$, which is taken to be {opposite to that} of the first:
\begin{equation}
\hat{H}=\hat{\sigma}_1^\mathrm{x}\hat{\sigma}_2^\mathrm{x}+h(\hat{\sigma}_1^\mathrm{z}-\hat{\sigma}_2^\mathrm{z}),\label{eq:spin1-1}
\end{equation}
In this case one can show that the steady-state of the system is independent of the magnetic field $h$. Accordingly, the first moment of the spin-flip statistic of spin $1$ is found to be simply
\begin{equation}
\kappa_1=\frac{2\gamma}{\gamma^2+4}.
\end{equation}
However even if the system steady-state is independent of the magnetic field, as is any of its observables, the second cumulant and the Fano factor do depend on it, yielding
\begin{equation}
\frac{\kappa_2+\kappa_1^2}{\kappa_1}=1+\frac{2}{16h^2+\gamma^2+4}+\frac{2(\gamma-5)}{\gamma^2+4}+\frac{32}{(\gamma^2+4)^2}.
\end{equation}
This example illustrates perfectly how the present iterative approach has the strong advantage to allow for the estimation of some parameters of the system that are usually not accessible. In \fref{fig:XZfano-1} we present a density plot, analogous to \fref{fig:XZfano}, representing the Fano factor as a function of the magnetic field $h$ and spin flip rate $\gamma$.

Contrary to what we had previously, here the system output can present antibunching, i.e., $(\kappa_2+\kappa_1^2)/\kappa_1<1$ occurs for certain regions in the parameter space; more specifically we have that $1/2\leq(\kappa_2+\kappa_1^2)/\kappa_1\leq9/8$. The maximal antibunching obtained corresponds to an infinite magnetic field $h$ and $\gamma$ tending to $0$, converging to a Fano factor equal to $1/2$. The maximal bunching corresponds to $\gamma\to2(2\pm\sqrt{3})$ while the magnetic field goes to $0$. As is clearly visible in \fref{fig:XZfano-1}, for $\gamma=2$ we have the special feature that there is always antibunching whatever the value of the magnetic field $h$.

We have used these two examples to illustrate that moments of the statistics of excitation-exchange can be efficiently obtained analytically, and also used to infer system parameters. Furthermore, we have shown explicitly that the information so obtained is greater than can be obtained simply by evaluating expectation values of the system. Our proposed approach therefore enables the possibility to unravel exactly and analytically the cumulants of the exchange statistics. However, as we have already discussed, the large-deviation approach encompasses more than just full-counting statistics. In order to illustrate this, let us next consider two interacting spins that also undergo global damping.

\subsection{Two globally damped spins-$1/2$}
We consider a system of two spins-$1/2$ evolving under the action of the Hamiltonian
\begin{equation}
\hat{H}=\hat{\sigma}_1^\mathrm{x}\hat{\sigma}_2^\mathrm{x},
\end{equation}
as well as under a single damping channel given by the Liouville operator $\hat{L}=\hat{\sigma}_1^+\hat{\sigma}_2^+$ and an associated rate $\gamma$. Under these conditions, the counting process we shall investigate reflects the global spin-flip {behaviour}, and we find the following expression for the first cumulant:
\begin{equation}
\kappa_1=\frac{4\gamma}{\gamma^2+8}[\rho_{\uparrow\uparrow}(0)+\rho_{\downarrow\downarrow}(0)],\label{eq:moment1}
\end{equation}
where $\rho_{\uparrow\uparrow}(t)=\langle\uparrow\uparrow\vert\hat{\rho}(t)\vert\uparrow\uparrow\rangle$ and $\rho_{\downarrow\downarrow}(t)=\langle\downarrow\downarrow\vert\hat{\rho}(t)\vert\downarrow\downarrow\rangle$ are, respectively, the probability to have the two spins up and the two spins down at a time $t$. What we see here is that the long-time limit {behaviour} of the exchange statistics relies strongly on the initial configuration of the system. For all initial states having $\rho_{\uparrow\uparrow}(0)+\rho_{\downarrow\downarrow}(0)=0$, we will have a quiet system, which will experience zero global spin flips. This is a rather intuitive result considering the form of the Liouville operator coupling the system to the environment, and leads to a different {behaviour} of the system depending on the initial conditions.

\begin{figure}
\centering
\includegraphics[scale=0.65]{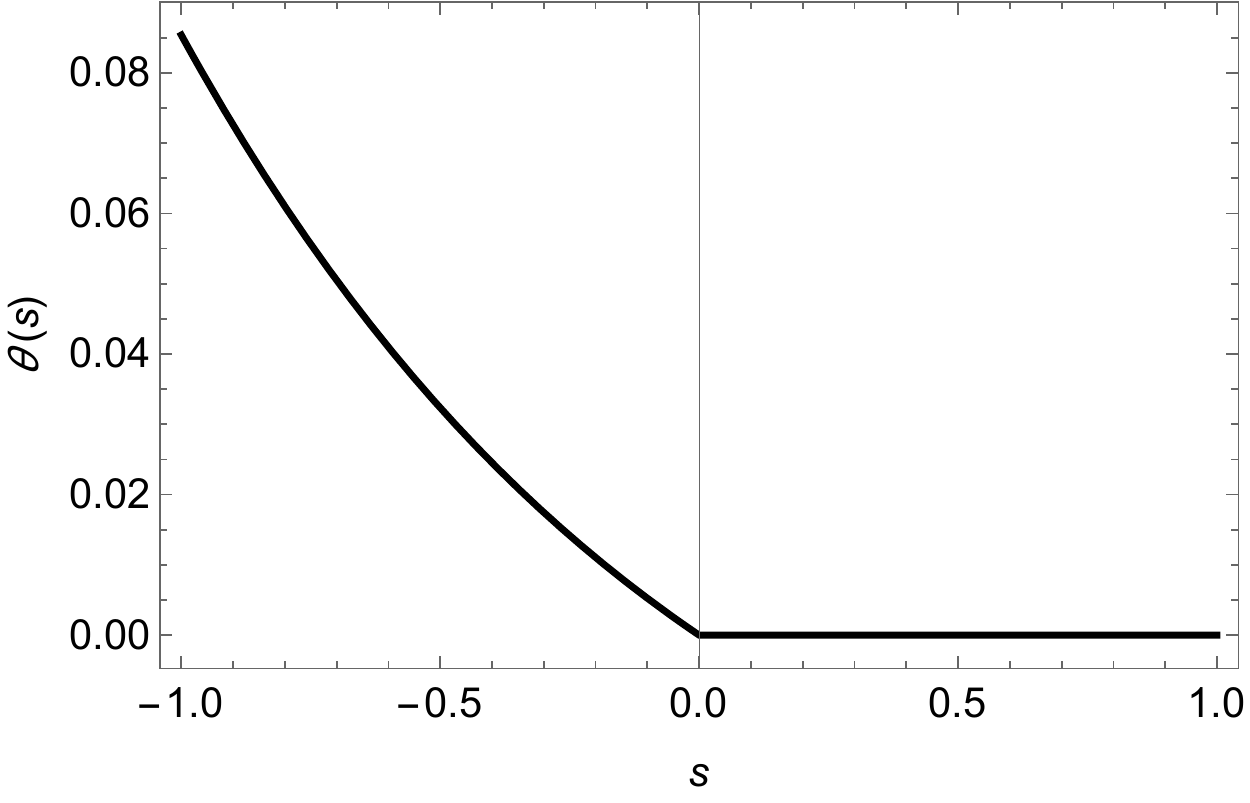}
\caption{Large deviation function $\theta(s)$ as defined in \eref{eq:theta} for a global spin-flip rate $\gamma=0.1$.\label{fig:xxglobal}}
\end{figure}

Other approaches for accessing directly the large-deviation function $\theta(s)$ are applicable in the present case because of the small size of the Liouville space that the system resides in; however, such approaches are actually less insightful than the one we presented above. For example, through direct resolution of the long-time {behaviour} of the biased matrix given by the master equation~(\ref{eq:ldf-integral}), we can obtain the following expression for the large-deviation function:
\begin{equation}
\theta(s)=
\left\{\begin{array}{cc}\frac{3^{2/3}\gamma^2(\gamma^2-16)+\sqrt[3]{3}F_{s}(\gamma)^2-3\gamma^2F_{s}(\gamma)}{6\gamma F_{s}(\gamma)} & \mathrm{if}\quad s<0 \\
0 & \mathrm{otherwise}\end{array}\right. ,\label{eq:theta}
\end{equation}
with $F_{s}(\gamma)=\gamma\bigl[\sqrt{3}\sqrt{12^{3}\gamma^2e^{-2s}-(\gamma+4)^{3}(\gamma-4)^{3}}+72\gamma e^{-s}\bigr]^{1/3}$. From the definition of $\theta(s)$, we can directly deduce that this system undergoes a dynamical phase transition, because of the non-analyticity present in $\theta(s)$ at $s=0$; this is visible in \fref{fig:xxglobal}. The two distinct phases are related to $\theta(s\to0^{-})$ and $\theta(s\to0^{+})$, with the latter corresponding to a quiet scenario where no spin flips occur ($\kappa_{n+}=-\partial_{s}^{n}\theta(s)\vert_{0^{+}}=0$ for any $n$). The phase related to $\theta(s\to0^{-})$ presents a finite rate of spin flips, which agrees with \eref{eq:moment1} when we take $\rho_{\uparrow\uparrow}(0)+\rho_{\downarrow\downarrow}(0)=1$. Consequently direct resolution of the biased master equation does not{, in general,} allow to capture the origin of the existing dual dynamics {and its dependence on the initial conditions, even if this is obvious in the present case. The method we propose, through \eref{eq:moment1}, paints a more complete picture and describes clearly how the outcome will depend on the initial conditions}.\\

In this section we focused exclusively on small spin systems which evolve within a relatively small Liouville space. For these systems we saw some of the main advantages of the iterative method proposed here, namely (i)~allowing analytical determination of the first cumulants of the exchange statistics, and (ii)~revealing the origin of dynamical phase transitions. In the following section we will consider systems residing in an infinite Liouville space, which consequently cannot be resolved exactly through direct analysis of the biased density matrix and its dynamics as given in \eref{eq:ldf-integral}.

\section{Damped quantum harmonic oscillators}
In this section we will move away from systems residing in small Liouville spaces to ones residing in infinite spaces. Such systems pose challenges to traditional analyses, numerical or analytical, for obtaining the large-deviation function from the biased master equation. In previous work we showed how an exact analytical solution may be found in the case of a single harmonic oscillator~\cite{Pigeon2014a} and made steps towards obtaining efficient numerical solutions that may be used for generic networks of harmonic oscillators~\cite{Pigeon2015}. We stress once again that it is only in a few exceptional cases that the large-deviation function may be obtained analytically. In contrast, the present technique will allow to access not the large-deviation function itself, but analytical formulations for the first few cumulants. As a first example in an infinite Liouville space, let us consider a system composed of two squeezed thermal harmonic oscillators.

\subsection{Two squeezed thermal harmonic oscillators\label{sub:Two-thermal-squeezed}}
A system composed of two thermal squeezed harmonic oscillator modes is well-studied in the context of quantum optics, providing a simple paradigmatic system to study bipartite continuous variable states~\cite{Ferraro2005,Jeong2000}. Such a system is also easily experimentally realised, corresponding as it does to spontaneous parametric down-conversion~\cite{Collett1984,Mahboob2014,Purdy2014}. Most work done on this system so far has, however, focused on its internal dynamics. What we are interested in here, on the other hand, is the statistics of the output of the system, in a similar manner as our earlier work~\cite{Pigeon2015}.

We start by considering two quantum harmonic oscillators coupled through a squeezing-like interaction as described by the Hamiltonian
\begin{equation}
\hat{H}=\sum_{i=1}^2\omega\left(\hat{a}_i^\dagger\hat{a}_i+\frac{1}{2}\right)+g\left(\hat{a}_2\hat{a}_1+\mathrm{H.c.}\right),
\end{equation}
where $\hat{a}_i^{(\dagger)}$ ($i=1,2$) corresponds to the creation (annihilation) operator associated to the oscillator $i$. The two oscillators are assumed to have the same frequency and are damped independently by means of the Liouville operators $\hat{L}_i=\hat{a}_i$ and damping rates $\gamma_i$. We shall consider a counting process that keeps track of the number of excitations leaving the oscillator $1$ to its bath. Solving the biased master equation is not possible due to the infinite size of the Liouville space involved. Nonetheless, we showed in earlier work~\cite{Pigeon2015} that by taking into account the linear nature of the dynamics, originating from a Hamiltonian that is quadratic in the operators, making a Gaussian ansatz for the state of the system, and moving to a phase space representation, the exchange statistics can be explored in detail. In a nutshell, this framework reduces the problem to numerically solving an algebraic Riccati equation, for which efficient algorithms are available. However, here we are interested in accessing the analytical formulation of the first cumulants of the statistics. To this end, and as detailed in \ref{sec:a2}, we develop the same iterative method as before but this time in terms of Gaussian parameters that correspond to a phase space representation of the situation. There is one key difference between the method we apply here, specialised to Gaussian states, and the more general one in previous sections. Whereas before we proceeded by determining the long-time {behaviour} of a given order of the biased density matrix $\rho^{(n)}$ related to the master equation given in \eref{eq:ccdcd}, in this context it is the biased covariance matrix of the same order, $\mathbf{\Sigma}^{(n)}=\partial_s^n\mathbf{\Sigma}_s\vert_{s=0}$, that lies at the centre of our method. This biased covariance matrix is associated with a Lyapunov equation, \eref{eq:dddf}. Indeed, one computational advantage of this iterative method is that it requires the solution of a set of Lyapunov equations, one for each order of the biased covariance matrix, $\mathbf{\Sigma}^{(n)}$, rather than an algebraic Riccati equation for $\mathbf{\Sigma}_s$. This reduces significantly the complexity of the problem.

In the present case we have for the drift matrix
\begin{equation}
\mathbf{A}=
\left(\begin{array}{cccc}
  \frac{\gamma_1}{2} & -\omega & 0 & g\\
  \omega & \frac{\gamma_1}{2} & g & 0\\
  0 & g & \frac{\gamma_2}{2} & -\omega\\
  g & 0 & \omega & \frac{\gamma_2}{2}
\end{array}\right);
\end{equation}
the noise matrix is diagonal and has the form $\mathbf{D}=\bigotimes_{j=1}^2\frac{\gamma_i}{2}\mathbf{1}_{2\times2}$, where $\mathbf{1}_{2\times2}$ is the $2\times2$ identity matrix. Taking $\gamma:=\gamma_1=\gamma_2$ for simplicity, we obtain the first and second cumulants
\begin{eqnarray}
\kappa_1&=\frac{2g^2\gamma}{\gamma^2-4g^2+4\omega^2},\quad\mathrm{and}\\
\frac{\kappa_2+\kappa_1^2}{\kappa_1}&=1+g^2 \frac{4\gamma^2+(2\gamma+1)(\gamma^2-4g^2+4\omega^2)}{(\gamma^2-4g^2+4\omega^2)^2}.
\end{eqnarray}
We see that deriving exact analytic definition of the mean and variance of the number of excitations exchanged between mode $1$ and the environment becomes here a simple task, and may for example lead to estimation of key parameters such as $g$. Having access to such relations also allows connections to be made between the output statistics of a system and its state. For example, the degree of squeezing of the modes of such a system is a key parameter that is of prime importance. For the minimal quadrature of the mode's degrees of freedom we find in this case that~\cite{Dechoum2004}
\begin{equation}
S^{-}=\frac{1}{1+\sqrt{\frac{4g^2}{\gamma^2+4\omega^2}}},\label{eq:sq}
\end{equation}
which may be rewritten using the definition of the first cumulant as
\begin{equation}
S^{-}=\frac{1}{1+\sqrt{\frac{2\kappa_1}{\gamma+2\kappa_1}}}.\label{eq:sq-1}
\end{equation}
This connects the degree of squeezing directly to experimentally-measurable statistics.

This example again illustrates the possible advantage that one can get from the iterative method we have described, as applied to a harmonic oscillator network. The method itself is not restricted to linear dynamics as presented here, however. Indeed, it may also allow to draw rather fundamental statements about the {behaviour} of systems undergoing non-linear dynamics, based on the language of thermodynamics of trajectories. To illustrate this point, we shall now consider the dynamics of a harmonic oscillator with a Kerr non-linearity.

\subsection{A single mode with a Kerr non-linearity}
\begin{figure}
\centering
\includegraphics[scale=0.65]{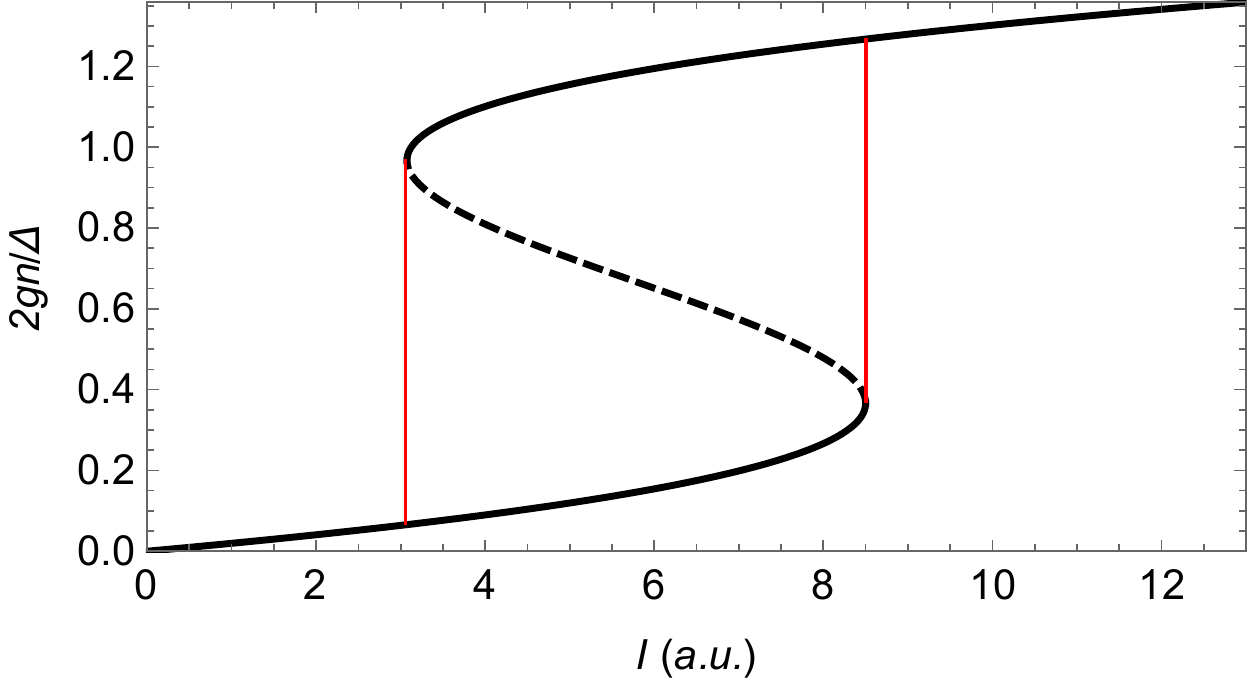}
\caption{Energy shift of the oscillator, $2gn$, due to non-linearity as a function of the driving intensity $I$ for $\gamma=\Delta/2$ and $g=\Delta/100$. The vertical red lines indicate the limit of the bistable region.\label{fig:kerr}}
\end{figure}

For this final example, we consider a quasi-resonantly driven oscillator possessing a Kerr non-linearity~\cite{Drummond1999,Imamoglu1997}. The dynamics of this system corresponds to various experiments, especially those conducted in the context of non-linear optical devices~\cite{Imoto1985,Kippenberg2004,Kirchmair2013}, and evolves under the Hamiltonian
\begin{equation}
\hat{H}=\hbar\omega\left(\hat{a}^\dagger\hat{a}+\frac{1}{2}\right)+g\hat{a}^\dagger\hat{a}^\dagger\hat{a}\hat{a}+F(e^{i\omega_\mathrm{p}t}\hat{a}^\dagger+e^{-i\omega_\mathrm{p}t}\hat{a}),
\end{equation}
where $F$ and $\omega_\mathrm{p}$ correspond respectively to the driving strength and frequency, and $g$ to the non-linear coupling. The operator $\hat{a}^{(\dagger)}$ is the creation (annihilation) operator of an excitation in the oscillator. We consider dissipation to take place to a single thermal bath, as described by the Liouville operator $\hat{L}=\hat{a}$. Using \eref{eq:11} and considering as a counting process the {net} number of {excitations} leaving the system to the environment, we solve for the stationary regime and find the first cumulant
\begin{equation}
\kappa_1=\gamma\langle\hat{n}\rangle\label{eq:11-1},
\end{equation}
where $\langle \hat{n} \rangle=\Tr\{\hat{a}^\dagger\hat{a}\rho(t\to\infty)\}$. In the Heisenberg picture we can derive an equation of motion for the operator $\hat{a}(t)$ and move to a mean-field approximation, where $\hat{a}(t)\to\Tr\{\hat{a}\rho(t)\} =\langle\hat{a}(t)\rangle=\alpha(t)e^{i\omega_\mathrm{p}t}$ in the appropriate rotating frame, yielding
\begin{equation}
i\partial_{t}\alpha=\left(\Delta-\frac{i}{2}\gamma-2g\vert\alpha\vert^2\right)\alpha+F,
\end{equation}
with $\Delta:=\omega-\omega_\mathrm{p}$. In steady-state we find
\begin{equation}
\left[(\Delta-2gn)^2+\frac{\gamma^2}{4}\right]n=I,\label{eq:la}
\end{equation}
where $n=\vert\alpha\vert^2\approx\langle\hat{n}\rangle$ is the mean field number of excitations and $I=\vert F\vert^2$ the driving
intensity. \eref{eq:la} links the number $n$ to the driving intensity, and corresponds to the hysteresis curve shown in \fref{fig:kerr}. We can refine our study by considering the stability of the mean field with respect to small fluctuations using the Hurwitz criterion~\cite{Drummond1999}, which guarantees stability for $n\leq n_-$ and $n\geq n_+$, where
\begin{equation}
n_\pm=\frac{2\Delta\pm\sqrt{\Delta^2-\frac{3\gamma^2}{4}}}{6g}.
\end{equation}
The unstable region, $n_-<n<n_+$, corresponds to the dashed part of the curve in \fref{fig:kerr}. From this plot we clearly see the rich dynamical {behaviour} of the system: For low enough driving the system remains in a linear phase, while for strong enough driving it will be in a non-linear phase. In between, due to the balance between non-linearity and dissipation, we find a bistable region where both solutions corresponding to the linear and non-linear phase are stable solution of the mean-field stationary state. More refined approaches can be used to extract more accurately the system {behaviour}, however the mean field already captures the main dynamical characteristics of the system.

As defined in \eref{eq:11-1}, the first cumulant of the output statistics is derived directly from the density of excitation of the system, $\langle\hat{n}\rangle$, which behaves differently in the three regimes previously mentioned, and which therefore depends strongly on the intensity of excitation $I$. The intermediate regime, corresponding to the bistable case, will consequently yield a situation where the first cumulant of the output of the system can take two distinct values $\kappa_1\approx\gamma n_\pm$ where
$n_\pm$ correspond to the density on the upper ($+$) and lower ($-$) curve in the bistable region. This is the signature of a first-order dynamical phase transition as explained in \sref{sub:Multi-stability-and-dynamical}. Based only on the first moment of the exchange statistics, we are able here to spot a dynamical phase transition in a well-known system. The link between between the thermodynamics of phase transitions and non-linear optical systems was previously considered~\cite{Carmichael2015,DeGiorgio1970},
but we see here how taking the point of view of thermodynamics of trajectories allows it to pass from a simple analogy to a correspondence.

\section{Conclusion}
In this article we proposed an efficient method to determine the statistics of the exchange of excitations between an open quantum system and its environment. This method is based on the formalism of thermodynamics of trajectories and has the key advantage of giving access to the cumulants, or moments, of these statistics with much less mathematical effort than that required to obtain the large-deviation function. The method we present is iterative, giving access to successive cumulants, and may be compared straightforwardly with experiment, where only the first few moments are usually accessible. We have shown that this method reveals more information on the nature of the dynamics undergone by the system than numerical evaluation of the large-deviation function, especially in scenarios where the system dynamics presents a dynamical phase transition. Lastly, we discussed and illustrated concretely the possibilities enabled by this method to cast new light on multi-stability of open quantum systems by interpreting such situations in the language of thermodynamics of trajectories.

\ack
The authors would like to thank Mauro Paternostro, Gabriele De Chiara{,} and Lorenzo Fusco for their fruitful discussions and support. 
This work was supported by the John Templeton Foundation (grant ID 43467), the Royal Commission for the Exhibition of 1851, COST Action MP1209 ``Thermodynamics in the quantum regime'' and the ANR ACHN C-Flight.


\appendix


\section{Formulation of the iterative approach in terms of expectation values of different orders of the biased density matrix $\rho^{(n)}$\label{sec:a1}}

In \sref{sub:Iterative} we presented an iterative method to obtain the first cumulants of the exchange statistics, formulated in terms of expectation values of the different orders of the biased density matrix $\rho^{(n)}=\partial_s^n\rho_{s}\vert_{s=0}$. Depending on the situation considered, and as summarised in \tref{tab:ssxs}, this formulation may be more efficient since it only consists of finding the long-time evolution of a number of expectation values. In the following, we define the expectation values that must be considered at each order, together with their respective evolutions, before we comment on possible advantages with respect to the density matrix formulation.

Considering the first cumulant $\kappa_1$, we do not actually need the full density matrix evolution $\rho(t)$ in principle. What is necessary is the expectation value associated to the channel $i$ as defined in \ref{eq:11}:
\begin{equation}
\kappa_1=\gamma_i\lim_{t\to\infty}\frac{1}{t}\int_0^t\rmd\tau\langle\hat{L}_i^\dagger\hat{L}_i(\tau)\rangle_0,
\end{equation}
where we define $\langle\hat{L}_i^\dagger\hat{L}_i(\tau)\rangle_0=\Tr\{\hat{L}_i^\dagger\hat{L}_i\rho(\tau)\}$. Based on the properties of the density matrix $\rho(t)$, the evolution of this expectation value can also be written as
\begin{equation}
\partial_{t}\langle\hat{L}_i^\dagger\hat{L}_i(t)\rangle_0=\Tr\{\hat{L}_i^\dagger\hat{L}_i\mathcal{W}[\rho(t)]\}.\label{eq:rho'me-1-1-1}
\end{equation}
The specific form taken by the right-hand side will depend on the system dynamics involved. Solving this equation in the long-time limit leads to the first cumulant $\kappa_1$. Notice that depending on the system dynamics in question, the right-hand side term of above equation might have a compact form, leading to straightforward results. This reduces, in principle, the difficulty of finding $\kappa_1$ with respect to the density matrix formation as presented in the main text.

Similarly, the second cumulant, defined as
\begin{equation}
\kappa_2=\gamma_i\lim_{t\to\infty}\frac{1}{t}\int_0^t\rmd\tau\bigl[\langle\hat{L}_i^\dagger\hat{L}_i(\tau)\rangle_0-2\langle\hat{L}_i^\dagger\hat{L}_i(\tau)\rangle_1\bigr],
\end{equation}
is connected to the evolution of the first-order biased matrix $\rho'${:}
\begin{eqnarray}
\langle\hat{L}_i^\dagger\hat{L}_i(\tau)\rangle_{n}&=\Tr\{\hat{L}_i^\dagger\hat{L}_i\rho^{(n)}(\tau)\}\\
&=\Tr\{\hat{L}_i^\dagger\hat{L}_i\partial_{s}^{n}\rho(\tau)\}\big\vert_{s=0}.
\end{eqnarray}
This matrix evolves according to \eref{eq:rho'me} which gives, for the evolution of the expectation values necessary to determine $\kappa_2$,
\begin{eqnarray}
\partial_{t}\langle\hat{L}_i^\dagger\hat{L}_i(t)\rangle_1=&\Tr\{\hat{L}_i^\dagger\hat{L}_i\mathcal{W}[\rho'(t)]\} \nonumber \\
&-\gamma_i\bigl[\langle\hat{L}_i^{\dagger2}\hat{L}_i^2(t)\rangle_0-\langle\hat{L}_i^\dagger\hat{L}_i(t)\rangle_0^2\bigr],\label{eq:rho'me-1-1}
\end{eqnarray}
where the last part of the right-hand side is the variance of the expectation values considered for the first cumulant, $\langle\hat{L}_i^\dagger\hat{L}_i(t)\rangle_0$. At this point we see that to calculate the second cumulant we need to be able to evaluate $\langle\hat{L}_i^\dagger\hat{L}_i\rangle_0$ as well as $\langle\hat{L}_i^{\dagger2}\hat{L}_i^2\rangle_0$, in order to obtain $\langle\hat{L}_i^\dagger\hat{L}_i\rangle_1$.

The third cumulant $\kappa_3$ is defined as
\begin{eqnarray}
\kappa_3=\gamma_i\lim_{t\to\infty}\frac{1}{t}\int_0^t\rmd\tau&[\langle\hat{L}_i^\dagger\hat{L}_i(\tau)\rangle_0-3\langle\hat{L}_i^\dagger\hat{L}_i(\tau)\rangle_1 \nonumber \\
&+3\langle\hat{L}_i^\dagger\hat{L}_i(\tau)\rangle_2].
\end{eqnarray}
{Given} that for $\kappa_1$ and $\kappa_2$, respectively, we previously evaluated $\langle\hat{L}_i^\dagger\hat{L}_i(\tau)\rangle_0$
and $\langle\hat{L}_i^\dagger\hat{L}_i(\tau)\rangle_1$, it remains to determine $\langle\hat{L}_i^\dagger\hat{L}_i(\tau)\rangle_2$,
which evolves according to
\begin{eqnarray}
\partial_{t}\langle\hat{L}_i^\dagger\hat{L}_i(t)\rangle_2&=\Tr\{\hat{L}_i^\dagger\hat{L}_i\mathcal{W}[\rho''(t)]\}\nonumber\\
&\qquad+\gamma_i\bigl(\langle\hat{L}_i^{\dagger2}\hat{L}_i^2(t)\rangle_0-\langle\hat{L}_i^\dagger\hat{L}_i(t)\rangle_0^2\bigr)\nonumber\\
&\qquad+4\gamma_i\langle\hat{L}_i^\dagger\hat{L}_i(t)\rangle_0\langle\hat{L}_i^\dagger\hat{L}_i(t)\rangle_1\nonumber\\
&\qquad-2\gamma_i\langle\hat{L}_i^{\dagger2}\hat{L}_i^2(t)\rangle_1
\end{eqnarray}
Here again we find that $\langle\hat{L}_i^\dagger\hat{L}_i\rangle_0$ appears. But other terms relying on $\rho'$ appear as well, such as $\langle\hat{L}_i^\dagger\hat{L}_i\rangle_1$. More importantly, other expectation values depending on $\rho$ and $\rho'$, such as $\langle\hat{L}_i^{\dagger2}\hat{L}_i^2\rangle_1$, are necessary in order to estimate the evolution of $\langle\hat{L}_i^\dagger\hat{L}_i\rangle_2$. However, it can be shown that the evaluation of $\langle\hat{L}_i^{\dagger2}\hat{L}_i^2\rangle_1$ relies on the one of $\langle\hat{L}_i^{\dagger3}\hat{L}_i^{3}\rangle_0$, which is obtained from $\rho$. Notice that for any $n\ge0$, each $\Tr\{\hat{L}_i^\dagger\hat{L}_i\mathcal{W}[\rho^{(n)}]\}$ will take the same form, and that $\Tr\{\rho^{(n)}\}=\delta_{n,0}$.

Generalising to the $n$\textsuperscript{th} cumulant, we have 
\begin{eqnarray}
\kappa_n=\gamma_i\sum_{j=0}^{n-1}{{n}\choose{j}}(-1)^j\lim_{t\to\infty}\frac{1}{t}\int_0^t\rmd\tau\langle\hat{L}_i^\dagger\hat{L}_i(\tau)\rangle_{j},
\end{eqnarray}
which requires us to evaluate the long-time evolution of all $\langle\hat{L}_i^\dagger\hat{L}_i(\tau)\rangle_{j}$ ($j<n$), which in turn obey
\begin{eqnarray}
\partial_{t}\langle\hat{L}_i^\dagger\hat{L}_i(t)\rangle_{n}=&\Tr\{\hat{L}_i^\dagger\hat{L}_i\mathcal{W}[\rho^{(n)}(t)]\}\nonumber \\
&\quad+\gamma_i\sum_{j=0}^{n-1}{{n}\choose{j}}\Biggl[(-1)^{n-j}\langle\hat{L}_i^{\dagger2}\hat{L}_i^2(t)\rangle_{j}\nonumber\\
&\quad-\sum_{a=0}^{j-1}{{j}\choose{a}}(-1)^{j-a}\langle\hat{L}_i^\dagger\hat{L}_i(t)\rangle_{a}\langle\hat{L}_i^\dagger\hat{L}_i(t)\rangle_{n-j}\Biggr].\label{eq:ccdcd-1}
\end{eqnarray}
To evaluate these expectation values, we therefore need to determine the evolution of expectation values such as $\langle\hat{L}_i^\dagger\hat{L}_i(t)\rangle_{a}$ and $\langle\hat{L}_i^{\dagger2}\hat{L}_i^2(t)\rangle_{a}$ for $a<n$. The first one will evolve accordingly to similar equation for $n\to a$, while the second needs to be handled with care. A similar equation to \eref{eq:ccdcd-1} can be obtained but will require higher powers of expectation values of lower order. For the $n$\textsuperscript{th} cumulant we therefore require $\langle\hat{L}_i^\dagger\hat{L}_i\rangle_{a}$ and $\langle\hat{L}_i^{\dagger2}\hat{L}_i^2\rangle_{a}$
both for $a<n$, $\langle\hat{L}_i^{\dagger3}\hat{L}_i^{3}\rangle_{a}$ for $a<n-1$, $\langle\hat{L}_i^{\dagger4}\hat{L}_i^{4}\rangle_{a}$ for $a<n-2$, etc., and $\langle\hat{L}_i^{\dagger n}\hat{L}_i^{n}\rangle_0$. In \tref{tab:ssxs} on the left-hand side are summarised the numbers of expectation values that must be evaluated to obtain the $n$\textsuperscript{th} cumulant. As previously discussed, the method is iterative, meaning that the determination of $\kappa_{n+1}$ when every other $\kappa_{n}$ is known requires only limited effort. More specifically, it will require to estimate the long-time {behaviour} of expectation values such as $\langle\hat{L}_i^\dagger\hat{L}_i(t)\rangle_{n}$, together with all $\langle\hat{L}_i^{\dagger2+a}\hat{L}_i^{2+a}(t)\rangle_{n-a}$ for $0\leq a<n+1$.

As is apparent from the foregoing discussion, this method is more cumbersome when working in terms of expectation values. However depending on the form taken by $\Tr\{\hat{L}_i^\dagger\hat{L}_i\mathcal{W}[\rho^{(n)}]\}$, this description can be computationally less demanding, especially in the case where the Liouville space associated to the system is large. For obtaining $\kappa_n$ in a matrix representation we need to solving the $n$ algebraic equations which can be rewritten as $n\times N^2$, differential equations where $N$ is the dimension of the Liouville space considered. For the same cumulant we will need to solve at least $\frac{1}{2}(n+1)n$ equations independently of the space dimension $N$. We remark, however, that depending on the form taken by $\Tr\{\hat{L}_i^\dagger\hat{L}_i\mathcal{W}[\rho^{(n)}]\}$, {it may be necessary to calculate the evolution of} some other expectation {values} coupled to this {one}, thus increasing the number of elements for which the evolution must be calculated. This independence of the number of expectation values necessary
to obtain the $n$\textsuperscript{th} cumulant on the size of the space $N$ is the chief advantage of working in this representation. For calculating the two first cumulants for a system evolving in a system space of dimension $10^{3}$, the expectation value description will require to solve at least three differential equations as opposed to {the} $3\times10^{3}$ {equations} required for the density matrix formulation. These differences are summarised in \tref{tab:ssxs}.

\begin{table*}
\centering
\caption{Description of the different quantities that must be evaluated in order to obtain the cumulants, using either the expectation-values formulation as in \eref{eq:ccdcd-1} (left side) or the density matrix formulation as in \eref{eq:ccdcd} (right-hand side). \label{tab:ssxs}}
\begin{tabular}{|c||c|c||c|c|}
\hline
\multirow{2}{*}{Cumulant} & Evolved expectation  & \multirow{2}{*}{\#} & \multirow{2}{*}{Evolved matrices necessary} & \multirow{2}{*}{\#}\tabularnewline
& values necessary (at least) &  &  & \tabularnewline
\hline
\hline
1 & $\langle\hat{L}_i^\dagger\hat{L}_i\rangle_0$ & 1 & $\rho$ & 1\tabularnewline
\hline
2 & $\langle\hat{L}_i^\dagger\hat{L}_i\rangle_0$, $\langle\hat{L}_i^{\dagger2}\hat{L}_i^2\rangle_0$
and $\langle\hat{L}_i^\dagger\hat{L}_i\rangle_1$ & 3 & $\rho$ and $\rho'$ & 2\tabularnewline
\hline
\multirow{2}{*}{3} & $\langle\hat{L}_i^\dagger\hat{L}_i\rangle_0$, $\langle\hat{L}_i^{\dagger2}\hat{L}_i^2\rangle_0$,
$\langle\hat{L}_i^{\dagger3}\hat{L}_i^{3}\rangle_0$ & \multirow{2}{*}{6} & \multirow{2}{*}{$\rho$, $\rho'$ and $\rho''$} & \multirow{2}{*}{3}\tabularnewline
& $\langle\hat{L}_i^\dagger\hat{L}_i\rangle_1$, $\langle\hat{L}_i^{\dagger2}\hat{L}_i^2\rangle_1$
and $\langle\hat{L}_i^\dagger\hat{L}_i\rangle_2$ &  &  & \tabularnewline
\hline
$\vdots$ & $\vdots$ & $\vdots$ & $\vdots$ & $\vdots$\tabularnewline
\hline
\multirow{2}{*}{$n$ (even)} & Any $\langle\hat{L}_i^{\dagger a}\hat{L}_i^{a}\rangle_{b}$ for
$b<n$ and  & \multirow{2}{*}{$\frac{n}{2}(n+1)$} & \multirow{2}{*}{Any $\rho^{(b)}$ for $b<n$} & \multirow{2}{*}{$n$}\tabularnewline
& $a\le n-b$ &  &  & \tabularnewline
\hline
\end{tabular}
\end{table*}


\section{Iterative approach for linear dynamics of harmonic oscillators\label{sec:a2}}
In this appendix, we will restrict ourselves to considering a network of quantum harmonic oscillators evolving linearly without any driving. Under those conditions, as we developed in Ref.~\cite{Pigeon2015}, the large-deviation function $\theta(s)$ related to the net number of excitation exchange between oscillator $i$ and its corresponding bath, $K:=K_{i-}-K_{i+}$, where $-$ refers to outgoing excitations and $+$ to incoming ones, can be written as
\begin{equation}
\theta(s)=\frac{1}{2}\Tr\{\mathbf{F}_{+}\mathbf{\tilde{\Sigma}}_{s}-\mathbf{F}_{-}\},\label{eq:mvtheta-1-2}
\end{equation}
where the matrix $\mathbf{F}_\pm$ is derived from the counting process $K$ related to the oscillator $i$,
\begin{equation}
\mathbf{F}_\pm=\bigotimes_{j=1}^{N}\delta_{i,j}
\left(\begin{array}{cc}
  f_{j\pm}(s) & 0\\
  0 & f_{j\pm}(s)
\end{array}\right),
\end{equation}
with $f_{i\pm}(s)=\gamma_i(e^{-s}-1)\pm\bar{\gamma}_i(e^{s}-1)$, with $\gamma_i$ and $\bar{\gamma}_i$ being rates defined in Ref.~\cite{Pigeon2015}, and $\mathbf{\tilde{\Sigma}}_{s}$ the stationary solution of the
biased covariance matrix according to the algebraic Riccati equation
\begin{equation}
0=\left(\mathbf{A}-\mathbf{F}_{-}\right)\tilde{\mathbf{\Sigma}}_{s}+\mathbf{\tilde{\mathbf{\Sigma}}}_{s}\left(\mathbf{A}-\mathbf{F}_{-}\right)^\mathrm{T}+\mathbf{\tilde{\mathbf{\Sigma}}}_{s}\mathbf{F}_{+}\mathbf{\tilde{\mathbf{\Sigma}}}_{s} \nonumber+\mathbf{F}_{+}-2\mathbf{D}.\label{eq:sigma}
\end{equation}
The matrices $\mathbf{A}$ and $\mathbf{D}$, respectively, are the drift and diffusion matrix related to the dynamics of the unbiased system in the phase space representation.

Expanding around $s=0$ we find
\begin{equation}
\tilde{\mathbf{\Sigma}}_{s}=\sum_{i=0}^{\infty}\frac{s^{i}}{i!}\mathbf{\Sigma}^{(i)}
\end{equation}
and
\begin{equation}
\mathbf{F}_\pm^{(n)}=\bigotimes_{j=1}^{N}\delta_{i,j}
\left(\begin{array}{cc}
f_{j\pm}^{(n)} & 0\\
0 & f_{j\pm}^{(n)}
\end{array}\right),
\end{equation}
with $f_{i\pm}^{(n)}=(-1)^{n}\gamma_i\pm\bar{\gamma}_i$ for $n>0$, otherwise $f_{i\pm}^{(0)}=0$, and where $\mathbf{\Sigma}^{(n)}=\partial_{s}^{(n)}\mathbf{\Sigma}_{s}\vert_0$ is $n$\textsuperscript{th} derivative with respect to $s$. Given this definition
the $n$\textsuperscript{th} cumulant can be written as
\begin{equation}
\kappa_{n}=\frac{1}{2}(-1)^{n}\sum_{i=0}^{n-1}{{n-1}\choose{i}}\Tr\bigl\{\mathbf{F}_{+}^{(n-i-1)}\mathbf{\Sigma}^{(i)}-\delta_{n-1,i}\mathbf{F}_{-}^{(n-1)}\bigr\}.
\end{equation}

As for the density matrix approach we see that here the $n$\textsuperscript{th} cumulant depends on lower orders of the biased density covariance matrix, $\mathbf{\Sigma}^{(i)}$, i.e., for $i<n$. Those orders of the biased covariance matrix will, in steady-state, be solutions of the following algebraic equation:
\begin{equation}
\mathbf{0}=\mathbf{A}\mathbf{\Sigma}^{(i)}+\mathbf{\Sigma}^{(i)}\mathbf{A}^\mathrm{T}+\mathbf{N}_i,\label{eq:dddf}
\end{equation}
with
\begin{eqnarray}
\mathbf{N}_{n}=&-\sum_{i=0}^{n-1}{{n-1}\choose{i}}\bigl(\mathbf{F}_{-}^{(n-i-1)}\mathbf{\Sigma}^{(i)}+\mathbf{\Sigma}^{(i)}\mathbf{F}_{-}^{(n-i-1)T}\bigr)\nonumber\\
&+\sum_{i,j,k=0}^{n-1}{{n-1}\choose{i,j,k}}\bigl(\mathbf{\Sigma}^{(i)}\mathbf{F}_{+}^{(j)}\mathbf{\Sigma}^{(k)}\bigr)+\mathbf{F}_{+}^{(n)}-2\delta_{n,1}\mathbf{D}
\end{eqnarray}
We see here that we have the phase-space equivalent {of} \eref{eq:ccdcd}. The matrix $\mathbf{N}_{n}$ represents a noise matrix related to the biasing procedure. \eref{eq:dddf} is known as a Lyapunov equation and can be solved using numerical methods but, as {reported} in \sref{sub:Two-thermal-squeezed}{,} may be solved exactly for some system. This, finally, gives access to the different cumulants of the exchange statistics related to the counting process $K$. Considering the two first {cumulants} we have that
\begin{eqnarray}
\kappa_1=&-\frac{1}{2}\Tr\{\mathbf{F}_{+}'\mathbf{\Sigma}-\mathbf{F}_{-}'\},\quad\mathrm{and}\\
\kappa_2=&\frac{1}{2}\Tr\{ 2\mathbf{F}_{+}'\mathbf{\Sigma}'+\mathbf{F}_{+}''\mathbf{\Sigma}-\mathbf{F}_{-}''\},
\end{eqnarray}
with the zeroth- and first-order biased covariant matrices being solutions of
\begin{eqnarray}
\mathbf{0}=&\mathbf{A}\mathbf{\Sigma}+\mathbf{\Sigma}\mathbf{A}^\mathrm{T}-2\mathbf{D}\label{eq:dddf-1},\quad\mathrm{and}\\
\mathbf{0}=&\mathbf{A}\mathbf{\Sigma}'+\mathbf{\Sigma}'\mathbf{A}^\mathrm{T}-\mathbf{F}_{-}'\mathbf{\Sigma}-\mathbf{\Sigma}\mathbf{F}_{-}'+\mathbf{\Sigma}\mathbf{F}_{+}^{\prime T}\mathbf{\Sigma}+\mathbf{F}_{+}',
\end{eqnarray}
{respectively,} where the first equation corresponds to the stationary solution of the unbiased (bare) covariance matrix. The present approach can be extended to scenarios with time-dependent covariance matrices, such as ones involving time-dependent driving.


\section*{References}
\begin{thebibliography}{10}

\bibitem{Touchette2009}
Touchette H, 
{\it The large deviation approach to statistical mechanics},
2009 {\em Phys. Rep.} 
\href{http://dx.doi.org/10.1016/j.physrep.2009.05.002}{{\bf 478} 1--69}.

\bibitem{Garrahan2010}
Garrahan J P and I. Lesanovsky I,
{\it Thermodynamics of quantum jump trajectories},
2010 {\em Phys. Rev. Lett.}
\href{http://dx.doi.org/10.1103/PhysRevLett.104.160601}{ {\bf 104}, 160601}.

\bibitem{Lesanovsky2013}
Lesanovsky I, van Horssen M, Gu\c{t}\u{a} M and Garrahan J P,
{\it Characterization of Dynamical Phase Transitions in Quantum Jump Trajectories Beyond the Properties of the Stationary State},
2013 {\em Phys. Rev. Lett.}
\href{http://dx.doi.org/10.1103/PhysRevLett.110.150401}{{\bf 110} 150401}.

\bibitem{Derrida1998}
Derrida B and Lebowitz J L, 
{\it Exact Large Deviation Function in the Asymmetric Exclusion Process},
1998 {\em Phys. Rev. Lett.}
\href{http://dx.doi.org/10.1103/PhysRevLett.80.209}{ {\bf 80} 209}.

\bibitem{Pigeon2014a}
Pigeon S, Fusco L, Xuereb A, De Chiara G and Paternostro M, 
{\it Thermodynamics of trajectories of a quantum harmonic oscillator coupled to $N$ baths},
2015 {\em Phys. Rev. A}
\href{http://dx.doi.org/10.1103/PhysRevA.92.013844}{ {\bf 92} 013844}.

\bibitem{Saito2007}
Saito K and A. Dhar A, 
{\it Fluctuation theorem in quantum heat conduction},
2007 {\em Phys. Rev. Lett.}
\href{http://dx.doi.org/10.1103/PhysRevLett.99.180601}{ {\bf 99} 180601}.

\bibitem{Znidaric2014}
\c{Z}ˇnidari\c{c} M,
{\it Exact Large-Deviation Statistics for a Nonequilibrium Quantum Spin Chain},
2014 {\em Phys. Rev. Lett.}
\href{http://dx.doi.org/10.1103/PhysRevLett.112.040602}{ {\bf 112} 040602}.

\bibitem{Giardina2006}
Giardin\`a C, Kurchan J and Peliti L, 
{\it Direct Evaluation of Large-Deviation Functions},
2006 {\em Phys. Rev. Lett.}
\href{http://dx.doi.org/10.1103/PhysRevLett.96.120603}{ {\bf 96} 120603}.

\bibitem{Nemoto2014}
Nemoto T and Sasa S, 
{\it Computation of Large Deviation Statistics via Iterative Measurement-and-Feedback Procedure},
2014 {\em Phys. Rev. Lett.}
\href{http://dx.doi.org/10.1103/PhysRevLett.112.090602}{ {\bf 112} 090602}.

\bibitem{Lenstra1982}
Lenstra D, 
{\it Photon-number statistics in resonance fluorescence}, 
1982 {\em Phys. Rev. A} 
\href{http://dx.doi.org/10.1103/PhysRevA.26.3369}{ {\bf 26} 3369}.

\bibitem{Sanchez2007}
S\'anchez R, Platero G and Brandes T, 
{\it Resonance fluorescence in transport through quantum dots: Noise properties}, 
2007 {\em Phys. Rev. Lett.} 
\href{http://dx.doi.org/10.1103/PhysRevLett.98.146805}{ {\bf 98} 146805}.

\bibitem{Flindt2008}
Flindt C, Novotn\'y T, Braggio A, Sassetti M, and Jauho A P, 
{\it Counting statistics of non-Markovian quantum stochastic processes}, 
2008 {\em Phys. Rev. Lett.} 
\href{10.1103/PhysRevLett.100.150601}{ {\bf 100} 150601}.

\bibitem{Flindt2010}
Flindt C, Novotn\'y T, Braggio A and Jauho A P, 
{\it Counting statistics of transport through Coulomb blockade nanostructures: High-order cumulants and non-Markovian effects}
2010 {\em Phys. Rev. B}
\href{http://dx.doi.org/10.1103/PhysRevB.82.155407}{ {\bf 82} 155407}.

\bibitem{Ates2012}
Ates C, Olmos B, Garrahan J P and Lesanovsky I,  
{\it Dynamical phases and intermittency of the dissipative quantum Ising model},
2012 {\em Phys. Rev. A}
\href{http://dx.doi.org/10.1103/PhysRevA.85.043620}{ {\bf 85} 043620}.

\bibitem{Garrahan2007}
Garrahan J P, Jack R, Lecomte V, Pitard E, van Duijvendijk K and van Wijland F, 
{\it Dynamical First-Order Phase Transition in Kinetically Constrained Models of Glasses},
2007 {\em Phys. Rev. Lett.}
\href{http://dx.doi.org/10.1103/PhysRevLett.98.195702}{ {\bf 98}, 195702}.

\bibitem{Hickey2012}
Hickey J M, Genway S, Lesanovsky I and Garrahan J P, 
{\it Thermodynamics of quadrature trajectories in open quantum systems},
2012 {\em Phys. Rev. A}
\href{http://dx.doi.org/10.1103/PhysRevA.86.063824}{ {\bf 86} 063824}.

\bibitem{Hickey2013}
Hickey J M, Genway S, Lesanovsky I and Garrahan J P,
{\it Time-integrated observables as order parameters for full counting statistics transitions in closed quantum systems},
2013 {\em Phys. Rev. B}
\href{http://dx.doi.org/10.1103/PhysRevB.87.184303}{ {\bf 87} 184303}.

\bibitem{Pigeon2014b}
Pigeon S, Xuereb A, Lesanovsky I, Garrahan J P, De Chiara G and Paternostro M, 
{\it Dynamical symmetries and crossovers in a three-spin system with collective dissipation},
2015 {\em New J. Phys.}
\href{http://dx.doi.org/10.1088/1367-2630/17/1/015010}{ {\bf 17} 015010}.

\bibitem{Evans2002}
Evans D J and Searles D J, 
{\it The Fluctuation Theorem},
2002 {\em Adv. Phys.}
\href{http://dx.doi.org/10.1080/00018730210155133}{ {\bf 51} 1529}.

\bibitem{Pigeon2015}
Pigeon S, Fusco L, Xuereb A, De Chiara G and Paternostro M, 
{\it Thermodynamics of trajectories and local fluctuation theorems for harmonic quantum networks},
2015 {\em New J. Phys.}
\href{http://dx.doi.org/10.1088/1367-2630/18/1/013009}{ {\bf 18}, 013009}.

\bibitem{Lecomte2007}
Lecomte V, Appert-Rolland C and Wijland F,
{\it Thermodynamic Formalism for Systems with Markov Dynamics},
2007 {\em J. Stat. Phys.} 
\href{http://dx.doi.org/10.1007/s10955-006-9254-0}{{\bf 127} 51}

\bibitem{Flindt2013}
Flindt C and Garrahan J P, 
{\it Trajectory phase transitions, lee-yang zeros, and high-order cumulants in full counting statistics}, 
2013 {\em Phys. Rev. Lett.} 
\href{http://dx.doi.org/10.1103/PhysRevLett.110.050601}{ {\bf 110} 050601}.



\bibitem{Ferraro2005}
Ferraro A, Olivares S and Paris M G A. 
{\it Gaussian states in continuous variable quantum information},
2005 arXiv:\href{http://arxiv.org/abs/quant-ph/0503237}{quant-ph/0503237}.

\bibitem{Jeong2000}
Jeong H, Lee J and Kim M S,
{\it Dynamics of nonlocality for a two-mode squeezed state in a thermal environment},
2000 {\em Phys. Rev. A}
\href{http://dx.doi.org/10.1103/PhysRevA.61.052101}{ {\bf 61} 052101}.

\bibitem{Collett1984}
Collett M J and Gardiner C W,
{\it Squeezing of intracavity and traveling-wave light fields produced in parametric amplification},
1984 {\em Phys. Rev. A}
\href{http://dx.doi.org/10.1103/PhysRevA.30.1386}{ {\bf 30} 1386}.

\bibitem{Mahboob2014}
Mahboob I, Okamoto H, Onomitsu K and Yamaguchi H, 
{\it Two-mode thermal-noise squeezing in an electromechanical resonator},
2014 {\em Phys. Rev. Lett.}
\href{http://dx.doi.org/10.1103/PhysRevLett.113.167203}{ {\bf 113}, 167203}.

\bibitem{Purdy2014}
Purdy T P, Yu P L, Peterson R W, Kampel N S and Regal C A, 
{\it Strong optomechanical squeezing of light},
2014 {\em Phys. Rev. X}
\href{http://dx.doi.org/10.1103/PhysRevX.3.031012}{ {\bf 3} 031012}.

\bibitem{Dechoum2004}
Dechoum K, Drummond P D, Chaturvedi S and Reid M D, {\it Critical fluctuations and entanglement in the non- degenerate parametric oscillator},
2004 {\em Phys. Rev. A}
\href{http://dx.doi.org/10.1103/PhysRevA.70.053807}{ {\bf 70} 053807}.

\bibitem{Drummond1999}
Drummond P D and Walls D F, {\it Quantum theory of optical bistability. I. Nonlinear polarisability model},
1999 {\em J. Phys. A: Math. Gen.}
\href{http://dx.doi.org/10.1088/0305-4470/13/2/034}{ {\bf 13} 725}.

\bibitem{Imamoglu1997}
Imamo\={g}lu A, Schmidt H, Woods G and Deutsch M, {\it Strongly Interacting Photons in a Nonlinear Cavity},
1997 {\em Phys. Rev. Lett.}
\href{http://dx.doi.org/10.1103/PhysRevLett.79.1467}{ {\bf 79} 1467}.

\bibitem{Imoto1985}
Imoto N, Haus H A and Yamamoto Y, {\it Quantum nondemolition measurement of the photon number via the optical Kerr effect},
1985 {\em Phys. Rev. A}
\href{http://dx.doi.org/10.1103/PhysRevA.32.2287}{ {\bf 32} 2287}.

\bibitem{Kippenberg2004}
Kippenberg T J, Spillane S M and Vahala K J. {\it Kerr-nonlinearity optical parametric oscillation in an ultrahigh-Q toroid microcavity},
2004 {\em Phys. Rev. Lett.}
\href{http://dx.doi.org/10.1103/PhysRevLett.93.083904}{ {\bf 93} 18}.

\bibitem{Kirchmair2013}
Kirchmair G, Vlastakis B, Leghtas Z, Nigg S E, Paik H, Ginossar E, Mirrahimi M, Frunzio L, Girvin S M and Schoelkopf R J, {\it Observation of quantum state collapse and revival due to the single-photon Kerr effect},
2013 {\em Nature}
\href{http://dx.doi.org/10.1038/nature11902}{ {\bf 495} 205}.

\bibitem{Carmichael2015}
Carmichael H J, {\it Breakdown of photon blockade: A dissipative quantum phase transition in zero dimensions},
2015 {\em Phys. Rev. X}
\href{http://dx.doi.org/10.1103/PhysRevX.5.031028}{ {\bf 5} 031028}.

\bibitem{DeGiorgio1970}
DeGiorgio V and Scully M, {\it Analogy between the Laser Threshold Region and a Second-Order Phase Transition}, 
1970 {\em Phys. Rev. A}
\href{http://dx.doi.org/10.1103/PhysRevA.2.1170}{ {\bf 2} 1170}.

\end{thebibliography}
\end{document}